# "Resonance" phenomena in thermal diffusion processes in two-layer structures

Yu. G. Gurevich, G.N. Logvinov

*Centro de Investigación y de Estudios Avanzados del Instituto Politécnico Nacional.*

*Apartado Postal 1470, C.P. 07000, México D.F. México.*

N. Muñoz Aguirre

*Oak Ridge National Laboratory, Bethel Valley Road, Oak Ridge, TN 37830, USA.*

L. Martínez Pérez

*Unidad Profesional Interdiciplinaria en Ingeniería y Tecnologías avanzadas del Instituto Politécnico Nacional, Av. I.P.N. No. 2580, Col. La Laguna Ticomán C.P. 07340, México D.F., México.*

February 6, 2002

**Abstract**

The dependence on chopper frequency of the effective thermal diffusivity and effective thermal conductivity in photoacoustic experiments is discussed. The theoretical model of a two-layer structure at rear-surface illumination in the high frequency limit is considered. It is shown that the effective thermal diffusivity presents "resonance" while the effective thermal conductivity sharply changes its magnitude and sign. Such "resonant" behavior strongly depends on the surface thermal conductivities associated with the interface thermal contacts.

One of the main problems of the heat transferring in multi-layer structures is the correct determination and measurement of the effective thermal conductivity and effective thermal diffusivity. The photoacoustic technique is widely used for the experimental study of these parameters [1], [2] due to its simplicity and high sensitivity. In order to obtain the photoacoustic signal, two different ways are generally used: closed photoacoustic cell (front-surface illumination) and open photoacoustic cell (rear-surface illumination) [3]. In accordance with the photoacoustic technique, we have suggested the adequate procedure for the calculation of the effective thermal conductivity and effective thermal diffusivity of two-layer samples [4]. The procedure is based on the idea that the photoacoustic signal or temperature response is only measured at one surface of the structure. Therefore, the two-layer structure can correctly be described by means of an imaginary sample whose volume is like a "black box" and it can be studied as an effective homogeneous one-layer system. Therefore, the requirement of the procedure is the equality of the temperatures at the front or rear-surface between the real two-layer structure and the effective one-layer sample. Then, the thermal parameters of the effective one-layer sample are the effective thermal parameters of the real two-layer structure. Such procedure is considered to be correct since the temperature is measured in the absolute Kelvin scale. Therefore, the "electrical and thermal analogy" [5], [6], generally used for the calculation of the effective thermal parameters, can be inadequate since it is based on the calculations of the temperature difference.

In a former work [4] we obtained general equations for the effective thermal parameters for both the low and high modulation frequency limits. However, the detailed analysis was only carried out for the low modulation frequency limit. In the present work we attempt to discuss the high modulation frequency limit within the frame of the "open photoacoustic cell" model. In this special case appears an unusual behavior of the effective thermal parameters connected with the following results:

1. As it was mentioned in [4], there is no single-valued determination of the effective thermal parameters in two-layer structures in photoacoustic experiments. The value of the effective thermal parameters depends on the point where the measurement is taken and the kind of measurement performed.

2. In the high frequency limit, the effective thermal conductivity and the effective thermal diffusivity are complex functions of the modulation frequency $\omega$ and they have a set of singularities [4].

3. As it will be shown later, the well known correlation [7]

$$\alpha_i = \frac{\kappa_i}{\rho_i c_i}, \qquad (1)$$

where $\alpha_i$ is the thermal diffusivity of $i^{th}$ layer, $\kappa_i$ is the thermal conductivity, $\rho_i$ is the density, $c_i$ is the specific heat, can not be used for the effective parameters in the general case.

In order to support the statements above, let us consider the simplest model of a two-layer structure. Let a slab to be made of the same material layers with thermal conductivity $\kappa$, thermal diffusivity $\alpha$, thickness $d$ and of unit cross section (See Figure 1). Moreover, let us consider that there exist an interface between the layers with surface thermal conductivity equal to $\eta$ [8]. Let us suppose that the incident laser beam is completely absorbed and transformed into heat on the surface $x=0$, in such way that the system is optically opaque. The opposite surface $x=2d$ is kept at the ambient temperature $T_o$ and the lateral sides are assumed to be adiabatic. Let us also consider all boundary conditions given in reference [4] be valid in this case and that the calculation of the effective thermal parameters is made at the surface $x = 2d-\varepsilon$, where $\varepsilon/2d << 1$. This theoretical procedure corresponds to the measurement of the effective parameters using the "open photoacoustic cell" method [4]. In the high frequency limit, the modulation frequency $\omega$ is large enough in such way that $\omega >> \omega_o$, where $\omega_o = 2\alpha/d^2$ is the characteristic frequency of the thermal diffusion process [7], [4]. Therefore, we use Eq. (22) of reference [4] to determine the effective thermal parameters. After some simple manipulations, we obtain two real equations to determine the effective thermal diffusivity $\alpha_{ef}$ and the effective thermal conductivity $\kappa_{ef}$,

$$\tan(2q_{ef}d) = \frac{\frac{\kappa q}{2\eta}(\sin(2qd)+\cos(2qd))+\sin(2qd)}{\frac{\kappa q}{2\eta}(\cos(2qd)-\sin(2qd))+\cos(2qd)}, \qquad (2)$$

$$\kappa_{ef} = \left(\frac{\kappa e^{2(q-q_{ef})d}}{\sin(2q_{ef}d)}\right)\left[\frac{\kappa q}{2\eta}(\sin(2qd)+\cos(2qd))+\sin(2qd)\right], \qquad (3)$$

where $q = \sqrt{\omega/2\alpha}$.

It is easy to see that $\alpha_{ef} = \alpha$ and $\kappa_{ef} = \kappa$ if $\eta >> \kappa q$. The last inequality holds under the condition $\omega >> \omega_o$, if $\xi = \eta d/\kappa >> 1$. We have got the natural result because the inequality $\xi >> 1$ means a perfect thermal contact between the slab's constituents [9], and the two-layer structure is reduced to the one-layer homogeneous sample. Let us now suppose that $\eta \ll \kappa q$, i.e. $\omega \gg \xi^2 \omega_o$. This condition allow us to write $\xi \leq 1$ or to assume the existence of an interface with finite surface thermal conductivity. In this case Eqs. (2), (3) reduce to

$$\tan(2q_{ef}d) = \frac{1 + \tan(2qd)}{1 - \tan(2qd)}, \tag{4}$$

$$\kappa_{ef} = \left(\frac{\kappa^2 q e^{2(q-q_{ef})d}}{2\eta}\right)\left[\frac{\sin(2qd) + \cos(2qd)}{\sin(2q_{ef}d)}\right]. \tag{5}$$

Plots of Eqs. (4) and (5) are respectively presented in Figures 2 and 3 for silicon layers of $d = 1 \times 10^{-4}$ m thickness each and taking $\eta = 10^3$ W/m$^2$°K (for silicon $\alpha = 0.89 \times 10^{-4}$ m$^2$/seg, $\kappa = 153$ W/m°K [10]). These plots show in detail the dependence of the effective thermal parameters on the modulation frequency. Moreover, the effective thermal diffusivity presents a "resonant" behavior at some frequencies as it can be observed in Fig. 2. Certainly, it is not the real resonance since it can only occur in real wave processes. On the other hand, the effective thermal conductivity sharply changes its magnitude and sign at these particular frequencies. This behavior is repeated as $\omega$ increases to infinity (see Fig. 3). At the same time, the Eqs. (4) and (5) show that the correlation (1) does not hold for the effective parameters even for the simplest two-layer structure. From this fact we can say that the effective thermal parameters lost their original meaning in comparison with the thermal parameters of each layer. The parameter $\alpha$ determines the thermal waves damping while $\alpha_{ef}$ does not have such meaning. Effective thermal parameters can be utilized for the ancillary aims, for example in order to determine the unknown thermal diffusivity of some layer if we know the thermal diffusivity of the other one and the effective value $\alpha_{ef}$. The same picture is true for the effective thermal conductivity, therefore it is not surprising that it can have negative values.

**Figure captions**

Figure 1: Two-layer structure.

Figure 2: "Resonance" behavior of the effective thermal diffusivity.

Figure 3: Effective thermal conductivity.

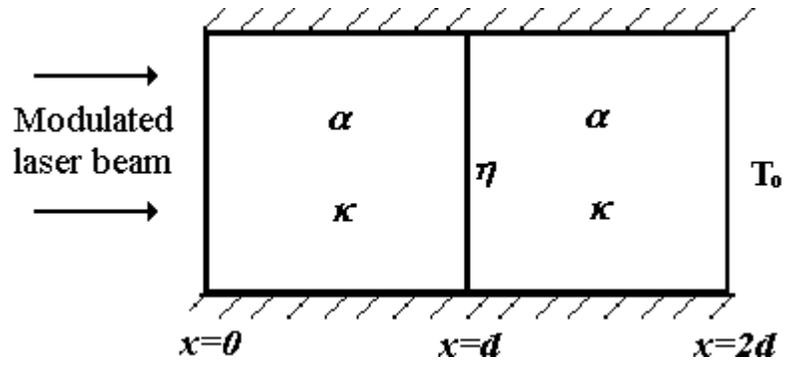

Figure 1

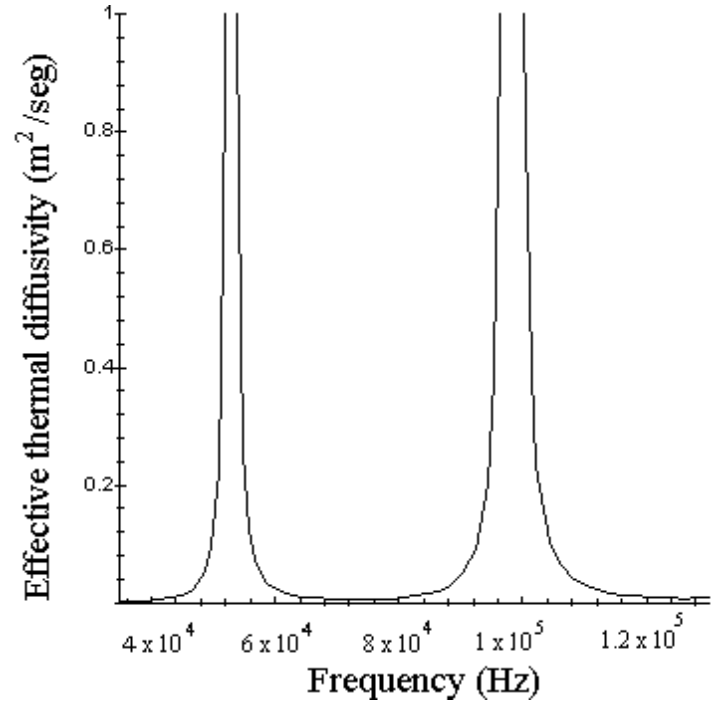

Figure 2

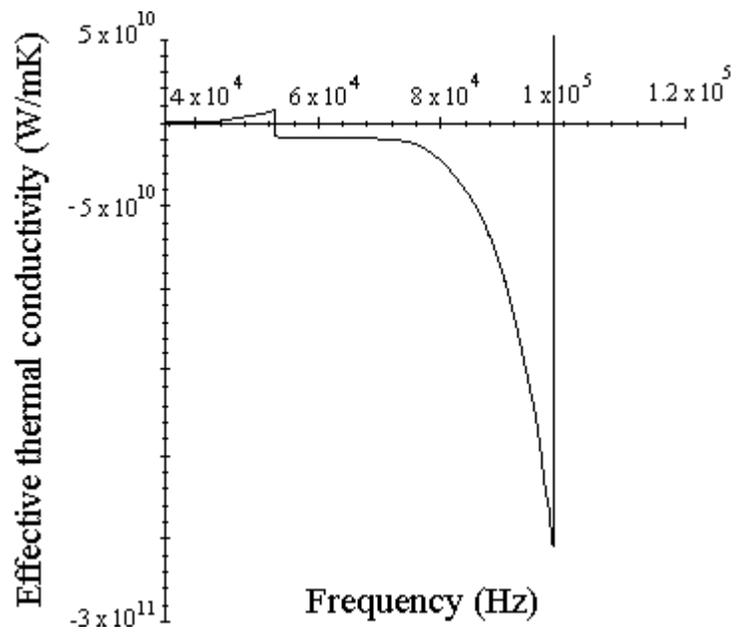

Figure 3

# ACKNOWLEDGMENT

N. Muñoz Aguirre would like to thank Consejo Nacional de Ciencia y Tecnología de México (CONACYT-México) for their Postdoctoral program and for providing financial support. This work was been supported by the Consejo Nacional de Ciencia y Tecnología (CONACYT), México.